\documentclass[cite]{epl}
\usepackage{amsmath,amssymb}
\usepackage[rightcaption]{sidecap}
\title{Time scales in shear banding of wormlike micelles}
\author{O.~Radulescu\inst{1} \and P.~D. Olmsted\inst{2} \and
  J.~P. Decruppe\inst{3} \and S.~Lerouge\inst{4} \and
  J.-F.~Berret\inst{5} \and G.~Porte\inst{5}}
\shortauthor{Radulescu \etal}
\institute{
  \inst{1} IRMAR - Universit\'e de Rennes1, Campus de Beaulieu, 35042,
  Rennes, France\\
  \inst{2}  Dept. of Physics and Astronomy, University
  of Leeds, Leeds LS2 9JT, UK\\
  \inst{3} LPLI - Universit\'e de Metz, Metz, France\\
  \inst{4} LLB - CEA-CNRS, Saclay, France \\
  \inst{5} GDPC - Universit\'e de Montpellier II,Montepellier,France }
 {Rheology, Structural
  and phase changes}  
\begin{document}
\maketitle
\begin{abstract}
  We show the existence of three well defined time scales in the
  dynamics of wormlike micelles after a step between two shear rates
  on the stress plateau. These time scales are compatible with the
  presence of a structured interface between bands of different
  viscosities and correspond to the isotropic band destabilization
  during the stress overshoot, reconstruction of the interface after
  the overshoot and travel of a fully formed interface. The last stage
  can be used to estimate a stress diffusion coefficient.
\end{abstract}
\section{Introduction}
Depending on the type and concentration of surfactant molecules and
added salt, solutions of surfactant wormlike micelles have shear
thinning or thickening behavior under shear flow. Unlike most fluids,
wormlike micelles often have non-analytic flow curves with
sharply-selected plateaus along which strain rate or stress may change
discontinuously.  In the well documented case of shear thinning
solutions the usual explanation of the constant stress plateau is
shear banding \cite{reh88,spenley,ber94}, \textit{i.e.}  a separation
of the material into bands of different viscosities, triggered by a
constitutive instability (such as an isotropic-to-nematic transition
\cite{ber94}). As shown recently \cite{rad99,olm00,rad00,lu99,yuan},
the stress selection and history independence of shear banding
can be explained using the inhomogeneities of the relevant mesoscopic
order parameter (polymer stress), \textit{i.e.} by incorporating
``diffusive'' terms in the constitutive equations. Order parameter
diffusion was introduced long time ago by van der Waals in the
so-called ``gradient theory'' of the gas-liquid interface \cite{vdw},
and is obligatory in phase field models for pattern formation.
Notwithstanding a few attempts to deal with inhomogeneous stresses
\cite{kar89} the same concept has not obtained full acceptance in the
rheological community.  While one might argue that diffusion terms are
negligibly small, these \textit{non-perturbative} terms resolve stress
selection even for infinitesimal values \cite{lu99}.  However, a small
diffusion coefficient should also imply a slow approach to steady
state; the main purpose of this letter is to demonstrate these long
time scales experimentally and relate them to simple model diffusive
behavior.

Shear banding involves spatial inhomogeneity and several temporal
stages.  Light polarization probes the local micellar orientation,
while rheology detects the molecular stress. In this work we shall
calculate the stress transients using a theoretical model. This will
be compared to birefringence measurements that are assumed to probe
the state of molecular orientation and hence stress (but see Ref.
\cite{call}) Rather than the typical start-up transient experiment, we
consider the simpler experiment of a step between two fixed values of
the shear rate in the banded regime. Small steps should induce less
drastic changes in the fluid while still remaining in the non-linear
regime, and hopefully yield more controllable results. The transient
features will be shown to be intimately related to the dynamics of the
interface between the bands.
\section{Experiments}
\begin{table}
  \begin{center}
  \caption{Surfactant systems used in this study; the surfactant was
    $0.3\,\mathrm{M}$ CTAB. $\tau$ and $G_0$, are the Maxwell
    relaxation time and plateau modulus, $\dot{\gamma}_I,\delta
    \dot{\gamma}$ are the start and width of the constant stress
    plateau, $(\tau_i)_{i=1,3}$ are the three time scales discussed in
    the text, and $\sigma^{\ast}$ is the plateau stress.}
  \label{t.1}
    \begin{largetabular}{ccccccccccc}
      No. & Salt & T [$^o$C] & $G_0$ [Pa] &$\tau$ [s] & $\tau_1$ [s] & $\tau_2$ [s] &
      $\tau_3$ [s] &
      $\tau \dot{\gamma}_I$ &
      $\tau \delta \dot{\gamma}$  & $\sigma^*/G_o$ \\\hline
      1 & 1.79M $NaNO_3$ & 30 &   232 &0.17  & 0.2 & 2.4 & 30.6 &0.85& 19  & 0.64  \\
      2 & 0.405M $NaNO_3$ & 30 &  238  &0.17  & 0.2 & 1.8 & 25 &1.27& 19 & 0.66  \\
      3 & 0.3M $KBr$ & 34 &  235 &0.16  & 0.2 & 3.7 & 9.3 & 1.12 &  80 & 0.66
    \end{largetabular}
  \end{center}
\end{table}
\begin{figure}
    \onefigure[scale=0.45]{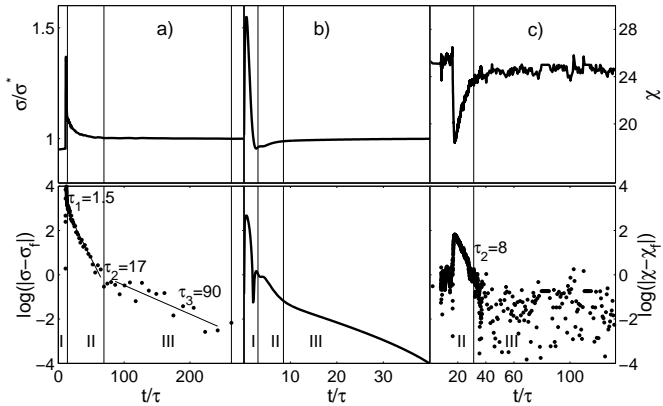}
\caption{Stages and
  characteristic times (in $\tau$ units): a) total stress during the
  step $\dot{\gamma}=10 \rightarrow 20\,\,s^{-1}$ for CTAB/KBr; b)
  simulated total stress during the step $\dot{\gamma}=1.4/\tau
  \rightarrow 2.8/\tau$ using the d-JS model; c) Extinction angle
  during the step $10 \rightarrow 20\,\,s^{-1}$ for CTAB/KBr. }
\label{f.1}
\end{figure}
The surfactant solutions used are summarized in Table~\ref{t.1}.
The stress response was measured using an RFS Rheometrics
Scientific controlled strain rate rheometer in Couette (radii
  $24.5, 25 \,\textrm{mm}$) and cone-plate geometries.  Linear response
is of the Maxwell type with almost identical relaxation times $\tau$
and moduli $G_0$ for the three solutions. In the non-linear regime the
coexistence plateau width $\delta\dot{\gamma}$ for solution $3$ is
larger than for solutions $1$ and $2$, while solutions $1$ and $2$
have a sloped stress plateau that roughly follows a power law $\sigma
\sim \dot{\gamma}^{\alpha}$, with $\alpha \lesssim 0.1$; a slope could
indicate concentration differences between coexisting states
\cite{schmitt95}.  Upon a step increase of the shear rate between
banded states the stress increases to a maximum, and then decreases
monotonically (or sometimes by a small undershoot and a monotonic
increase).
Three relaxation times ($\tau_1 \approx \tau \approx 10^{-1}\,s$,
$\tau_2 \approx 10 \tau \approx 1\,s$ and $\tau_3 \approx 100\tau
\approx 10s$) follow successively after the overshoot until steady
flow is reached (Fig.~\ref{f.1}a). The three time scales are well
separated and the result is reproducible for all solutions and
both cone-plate and Couette geometries. We filmed the step shear
rate experiment for the third solution in a Couette geometry with
a slightly larger gap (radii $24,25\,\textrm{mm}$) between crossed
polarizers to extract the average extinction angle $\chi$. The
kinetics of $\chi$ resemble that of the shear stress, displaying
time scales similar to the second and the third time scales found
in the rheology (Fig.~\ref{f.1}c).  The first, unresolvable, time
scale is shorter than the interval between successive video
frames.  The second time scale is well resolved, although shorter
by a factor of two than the equivalent time in the stress rheology
measurements. The third time scale is buried in noise and
regression does not provide significant results.
\section{Theory}
The momentum balance is $\rho\left(\partial_t +
  \vect{v}\!\cdot\!\boldsymbol{\nabla}\right)\vect{v} =
\boldsymbol{\nabla}\!\cdot\!\vect{T},$ where $\rho$ is the fluid
density and $\vect{v}$ is the velocity field. The stress tensor
$\vect{T}$ is given by $ \vect{T} = -p\,\vect{I} + 2\eta\vect{D} +
\boldsymbol{\Sigma}$, where the pressure $p$ is determined by
incompressibility ($\boldsymbol{\nabla}\!\cdot\!\vect{v}=0$), $\eta$
is the ``solvent'' viscosity, $\boldsymbol{\Sigma}$ is the ``polymer''
stress, and $\vect{D}$ is the symmetric part of the velocity gradient
tensor $(\boldsymbol{\nabla}\vect{v})_{\alpha\beta}\equiv
\partial_{\alpha}v_{\beta}$.  The non-Newtonian ``polymer''
viscoelastic stress $\vect{\Sigma}$ is assumed to obey the diffusive
Johnson-Segalman (d-JS) model \cite{olm00},
\begin{equation}
  \label{eq:JS}
\left(\partial_t +
    \vect{v}\!\cdot\!\boldsymbol{\nabla}\right)\boldsymbol{\Sigma}
  - \left(\boldsymbol{\Omega\Sigma} - \boldsymbol{\Sigma\,\Omega}\right)
  - a \left(\vect{D}\boldsymbol{\Sigma} + \boldsymbol{\Sigma}\vect{D}\right)
 = {\cal
    D}\nabla^2\boldsymbol{\Sigma} + 2 \mu \vect{D}/\tau -
  \boldsymbol{\Sigma}/\tau,
\end{equation}
where $\boldsymbol{\Omega}$ is the anti-symmetric part of
$\boldsymbol{\nabla}\vect{v}$, $\mu=G_o \tau$ is the ``polymer'' viscosity,
$\tau$ is a relaxation time, and ${\cal D }$ is the diffusion
coefficient.  The ``slip parameter'' $a$ (describing the
non-affinity of the deformation) is necessary to reproduce a
non-monotonic constitutive curve, and the added diffusion term was
shown to resolve stress selection \cite{olm00}.

The initial dynamics is governed by inertia; within a very short time
$\tau_M = \rho L^2 / \eta$ ($\tau_M= 10^{-4}\,\mathrm{s}$ for
$\eta/\rho = 0.01\, \textrm{m}^2\textrm{s}^{-1}$ and a gap
$L=1\,\mathrm{mm}$) momentum diffuses across the gap and the momentum
balance becomes $\vect{T}=const$.  The subsequent slower dynamics is
controlled by the viscoelastic response of the fluid.  In a planar
geometry with $\vect{v}=v(y)\hat{\vect{x}}$, our constitutive model
leads to a system of reaction-diffusion equations\footnote{Any
  non-monotonic differential constitutive model (\textit{e.g.}  Cates
  \cite{cat90}, Doi-Edwards \cite{doiedw,larson}) with diffusion terms
  leads to a similar equation set.}:
\begin{align}
  \frac{\partial S}{\partial t} & = {\cal D\/} \frac{\partial^2
    S}{\partial y^2} - \frac{S}{\tau} +
  C_S(\dot\gamma,S,W),&
  \frac{\partial W}{\partial t} & = {\cal D\/} \frac{\partial^2
    S}{\partial y^2} - \frac{W}{\tau} + C_W(\dot\gamma,S,W),\label{eq:system}
\end{align}
where $\dot\gamma$ is the shear rate, $S=\Sigma_{xy}$, and $W$ is
a combination of the polymer normal stresses, $\Sigma_{xx}$ and
$\Sigma_{yy}$. $S,W$ are the order parameters of the transition
($S$ is small in the nematic (N) band, and large in the isotropic
(I) band). They can diffuse across stream lines with diffusion
coefficient ${\cal D\/}$, and relax in the linear regime within
the linear (Maxwell) time $\tau$.  The non-linear reaction terms
$C_S=\dot\gamma (G_o-W)$ and $C_W=\dot\gamma S$ can be
straightforwardly derived from Eq.~(\ref{eq:JS})
\cite{rad99,rad00}.

The local momentum balance for the shear stress $\sigma=T_{xy}$ is
\begin{equation}
  \label{eq:stress}
\sigma = S + \epsilon G_o \tau \dot\gamma,
\end{equation}
where $\epsilon=\eta/\mu$. The dynamics of Eq.~\ref{eq:system} can be
understood with the aid of two (local) dynamical systems:
\begin{subequations}
\begin{align}
  \dot{S} & =  - S/\tau + C_S(\dot\gamma,S,W),&
  \dot{W} & =  - W/\tau + C_W(\dot\gamma,S,W);
  \label{eq:cshear}\\
  \dot{S}& =  - S/\tau + C_S((\sigma-S)/\epsilon G_0\tau,S,W),&
  \dot{W} & = - W/\tau + C_W((\sigma-S)/\epsilon G_0\tau,S,W)
  \label{eq:cstress}
\end{align}
\end{subequations}
where $\dot{S}\equiv\partial S/\partial t$.  System~(\ref{eq:cshear})
describes the dynamics along a streamline at prescribed shear rate; in
this case $\sigma$ changes proportionately to $S$ according to
Eq.~(\ref{eq:stress}).  System~(\ref{eq:cstress}) describes the
dynamics along a streamline at constant total stress $\sigma$.

\begin{figure}
  \centering{\includegraphics[scale=0.6]{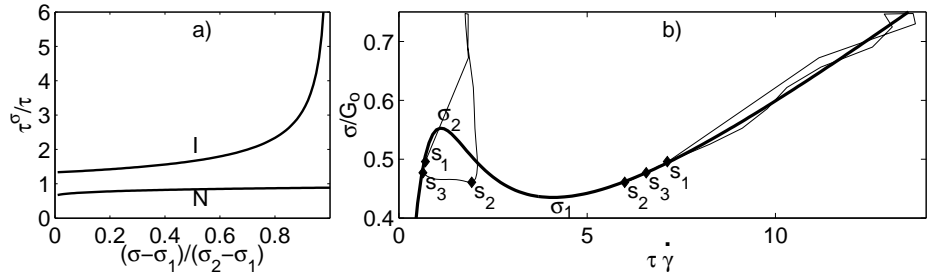}}
 \caption{a) Relaxation times to I and N attractors at constant stress,
   $\tau_I^{\sigma}, \tau_N^{\sigma}$. b) Thick line: theoretical
   flow curve (the negative slope,unstable branch should be replaced
   by the constant stress plateau).  Thin lines: trajectories of the
   coexisting bands near the walls. $\{s_i\}, i = 1,3$ are the
   starting points of the three stages discussed in the paper.
   }
 \label{f.2}
\end{figure}

The two dynamical systems have the same fixed points (since, for
homogeneous steady flow, $\dot\gamma$ and $\sigma$ are related by
Eq.~\ref{eq:stress}): stable fixed points (attractors)
representing the bands I and N, and an intermediate unstable
saddle fixed point. Coexistence of bands at common total stress is
possible only for $\sigma \in [\sigma_1(\epsilon),
\sigma_2(\epsilon)]$.  Linearizing systems~(\ref{eq:cshear})
and~(\ref{eq:cstress}) about the fixed points yields the dominant
relaxation times of the attractors, $\tau_{I}$ and $\tau_{N}$.
These are different for the two dynamical systems, denoted at constant
shear rate by $\tau_{I,N}^{\gamma}$ and at constant stress by
$\tau_{I,N}^{\sigma}$. For the JS model
$\tau_{I}^{\gamma}=\tau_{N}^{\gamma}=\tau$ for all $\dot{\gamma}$.
$\tau_{N}^{\sigma}$ is close to $\tau$, while $\tau_{I}^{\sigma}$ is
larger than $\tau_{N}^{\sigma}$ and diverges as $\sigma \rightarrow
\sigma_2(\epsilon)$ (Fig.~\ref{f.2}a). This divergence is consistent
with Ref.~\cite{gra97}: controlled stress experiments have much longer
relaxation times on the metastable extension of the high viscosity
branch above the constant stress plateau.

Consider an initial banded steady state, with average shear rate
$\langle\dot\gamma\rangle$.  Suddenly increasing the average shear
rate to $\dot\gamma_2 > \langle\dot\gamma\rangle$ produces a stress
overshoot because the amount of high viscosity I band is too large;
both I and N bands then become unstable, and the stress can decrease
by producing more of the low viscosity N material. Numerical
simulation (Figs.\ref{f.1}, \ref{f.3}) shows that this occurs in three
stages:

\begin{figure}
  \onefigure[scale=0.45]{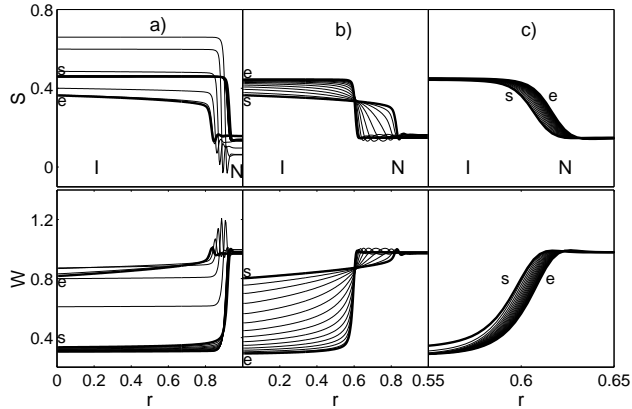}
 \caption{Simulation
   of the order parameter profiles for the three stages (s: start, e:
   end): a) destabilization ($0<t<t_1$); b) reconstruction
   $(t_1<t<t_2)$; c) travel ($t_2<t<\infty$). }
 \label{f.3}
\end{figure}
\textit{1. Band I destabilization}--- During this stage the I band tries
a direct passage toward the nematic band N, Fig.\ref{f.3}a).
Complete transformation is forbidden by the average shear rate
constraint and the I band stops before reaching the basin of
attraction of the steady N band.
A representation of the subsequent kinetics in the $(\dot\gamma,
\sigma)$ plane (Fig.~\ref{f.2}b) shows that \textit{the N band
almost
  follows the steady flow curve, while the I band evolves at constant
  shear rate}. Thus, the characteristic time $\tau_1 \approx
\tau_I^{\gamma} \approx \tau$ is controlled by the I band dynamics.
The total shear stress at the end of this stage depends on the final
position of the interface and is sensitive to the details of the
constitutive model.  If this value is below the plateau one finds an
undershoot (as in the numerical simulation Fig.~\ref{f.1}b), and
otherwise the subsequent stress evolution is monotonic
(Fig.~\ref{f.1}a).  Although we did not succeed in reproducing
monotonic evolution using the JS model, as long the I band evolves at
a constant shear rate, the magnitude of the characteristic time should
not be affected by the presence or absence of an undershoot.

\begin{figure}
  \begin{center}
  \includegraphics[scale=1.0]{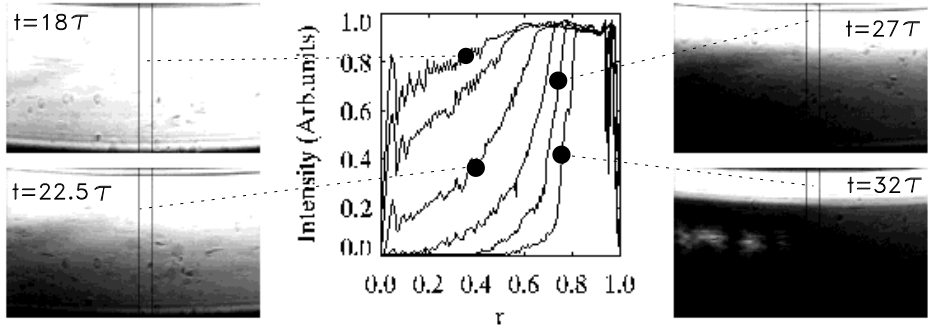}
  \end{center}
\caption{CTAB/KBR: $\dot{\gamma} = 10\rightarrow30 s^{-1}$ jump:
  birefringence images and profiles (averaged between the two vertical
  lines on the film) corresponding to the second time scale, showing
  the interface reconstruction. The moving (inner) cylinder is at
  $r=1.0$ and the fixed (outer) cylinder is at $r=0.0$.}
\label{f.4}
\end{figure}

\textit{2. Interface reconstruction}--- At the end of stage 1 the
interface separates an unsteady I band close to the unstable saddle
point from a nearly stable N band.  The part of the profile closer to
the I attractor will evolve toward this one, while the other part
approaches the N attractor. This reconstructs the interface in a more
advanced position, stabilizes the bands, and increases the contrast
between them. Interestingly, there is a spatial position at which $S$
and $W$ practically remain fixed at their saddle fixed point values.
During this stage $\langle\dot\gamma\rangle$ is constant (because it
is imposed) and $\langle S\rangle$ is almost constant (because of the
compensating evolutions of the two bands), so according to
Eq.~(\ref{eq:stress}) the total stress variation is small, and the
characteristic time (controlled by the I band) is $\tau_2 \approx
\tau_I^{\sigma}$. This time exceeds the linear viscoelastic time
$\tau$ (see Fig.~\ref{f.2}), and depends on how close the stress at
the end of stage 1 is to the spinodal limit $\sigma_2(\epsilon)$, and
on the quantitative details of the curve in Fig.~\ref{f.2}, all of
which are sensitive to the constitutive model. The analysis suggested
by the numerical experiment is confirmed by the birefringence
measurements. The sequence of images in Fig.~\ref{f.4} show the gap of
the Couette cell filmed between crossed polarizers during stage 2.
Although we can not quantitatively compare Figs.~\ref{f.3}b) and
\ref{f.4} (the relation between the transmitted intensity and the
order parameter is unknown and sure to be non-linear), the sharpening
of the contrast corresponding to the interface reconstruction is
visible. The difference between the characteristic times for the
extinction angle and rheology (Fig.\ref{f.1}) could be due to the
different Couette cell gap widths.

\textit{3. Interface travel}--- The instability and reconstruction of
the interface in the first two stages is ensured by the reaction terms
of the Eq.~(\ref{eq:system}), ending when a sharp interface between
stable bands is fully formed. This interface could have a non-zero
velocity if it forms in at a position corresponding to a stress value
above or below the plateau stress $\sigma^{\ast}$.  ``Front
propagation'' over the small distance toward the final equilibrium
position is then controlled by ${\cal D\/}$ (this distance is too
small to observe by birefringence). Because of the undershoot in the
numerical simulation the sign of the displacement during stage 3 is
opposite to the one in the first stages (Fig.\ref{f.3}c). The
characteristic time $\tau_3$ for this stage follows from the
velocity $c$ of the sharp interface close to steady state which is
independent of the presence or absence of the undershoot \cite{rad00}
(see below).

Let us consider a single sharp interface, at a position $r$ inside the
gap.  At imposed shear rate, the lever rule
$\langle\dot{\gamma}\rangle=\frac{r}{L}\dot{\gamma}^N(\sigma)+
(1-\frac{r}{L})\dot{\gamma}^I(\sigma)$ relates $\sigma$ and $r$ and
leads to
\begin{equation}
\left( \frac{\partial \sigma }{\partial
r}\right)_{\langle\dot{\gamma}\rangle}= \frac{-\eta_I
\delta \dot\gamma}{L[1+(\langle\dot\gamma\rangle-\dot\gamma_I)/\dot\gamma_I]},
\label{eq:drivingf}
\end{equation}
where $\delta\dot\gamma=\dot{\gamma}_N-\dot{\gamma}_I$ is the
width of the plateau, and $\eta_I= \left.\partial\sigma/ \partial
\dot{\gamma} \right|_{\gamma_I}$.  We consider
$\langle\dot\gamma\rangle-\gamma_I\ll \delta\dot\gamma$,
$\delta\dot{\gamma}\gg\dot{\gamma}_I$ (as in the experiments) so that 
$\eta^I \dot{\gamma}^I \approx \eta^N \delta\dot\gamma$
(true for piecewise linear flow curves and obeyed well by the JS
model).

We showed previously that the velocity $c$ of the interface is a
function of the total shear stress $\sigma$ only, and that $c=0$
when $\sigma=\sigma^*$ \cite{rad99,rad00} which via the lever rule
corresponds to a unique stable interface position $r_{\ast}$.
Thus, close to this position $r_*$ the equation of motion of the
interface is:

\begin{equation}
  \label{eq:2}
\frac{dr}{dt}=c(\sigma)=\frac{d c}{d \sigma}\left( \frac{\partial
    \sigma }{\partial r}\right)_{\langle\dot{\gamma}\rangle} (r-r^*).
\end{equation}

Using Eqs.~(\ref{eq:drivingf},\ref{eq:2}), and the derivative
$\left.\frac{d \sigma}{d c}\right|_{\sigma=\sigma^{\ast}} \equiv K G_o
\sqrt{\frac{\tau}{{\cal D\/}}}$, we find the solution $r-r^* =
(r(0)-r^*)e^{-t/\tau_3}$, with characteristic time
\begin{equation}
\tau_3 = K \frac{L}{\sqrt{{\cal D\/}
\tau}}\frac{1+[\langle\dot\gamma\rangle-\dot\gamma_I]/\dot\gamma_I}{\delta
\dot\gamma}, \label{eq:time3}
\end{equation}
where the dimensionless constant $K$ depends on the particular
constitutive model.

Eq.~(\ref{eq:time3}) implies that a fully formed interface
equilibrates faster in systems with larger plateaus $\delta \dot
\gamma$, such as CTAB/KBr. In such cases, Eq.~(\ref{eq:drivingf})
implies larger stress variations and thus larger interface
accelerations for the same position variation. This is compatible with
the shorter $\tau_3$ in Table~\ref{t.1}. For simplicity,
Eq.~(\ref{eq:drivingf}) was for a planar geometry; in cylindrical
Couette flow a slight correction (negligible for the thin gaps we
consider) leads to a smaller $\tau_3$.

Using the experimental value of $\tau_3$ and Eq.~\ref{eq:time3}, we
can estimate ${\cal D\/}$. In order to do this we need the value of
$K$. In the JS model, while $G_0$ and $\tau$ are measurable, the two
free parameters $\epsilon$ and $a$ determine the function $K$.
Nevertheless, $\tau G_o \delta\dot\gamma/\sigma^* = f_1(\epsilon)$,
$\dot\gamma_I/\delta\dot\gamma=f_2(\epsilon)$ and
$K/(\tau\delta\dot\gamma) = f_3(\epsilon)$ are functions of $\epsilon$
only, given to a good approximation ($>80\%$) by
$f_1(\epsilon)=\frac{4}{3\epsilon}\sqrt{\frac{1/8-\epsilon}{1/2-\epsilon}}$,
$f_2(\epsilon)=
\frac{3\epsilon}{\sqrt{(1-8\epsilon)
    (1-2\epsilon)}+1-8\epsilon}$, $f_3(\epsilon) = \frac{3}{8}
\epsilon^2 f_1(\epsilon)$. From either $f_1$, or $f_2$ and
experimental data, one can estimate $\epsilon$ (the average of the two
values is given in Table~\ref{t.2}) and then $f_3$ gives K
(Table~\ref{t.2}). Thus we do not need the value of $a$.
Microscopically, we expect ${\cal D\/}=\zeta^2/\tau$ where $\zeta$ is
the stress correlation length.  In dilute solutions this should be the
micelle gyration radius \cite{kar89}, while in concentrated solutions
a reasonable candidate is the mesh size $\xi$, which can be estimated
from $G_o \sim kT/ \xi^3$. The results are presented in
Table~\ref{t.2}. The stress correlation length is of order the mesh
size, which is reasonable; however there is still no theory for such a
diffusive term in concentrated solutions.

To conclude, a general dynamical systems analysis of the d-JS model
provides plausible explanations for the observed time scales, and
consistent estimates of the stress diffusion coefficient ${\cal D\/}$.
Nevertheless, neither the d-JS model, nor
reptation-retraction-reaction models \cite{doiedw,cat90} can provide
perfect fits of transient stress curves in the nonlinear regime.
Also, it it conceivable that concentration differences between the
bands could influence the observed time scales. The difference of the
values of $\cal D$ between samples 1,2 and 3 (Table~\ref{t.2}) could
be a concentration effect, consistent with the different slopes of the
flow curves plateaus (tilted for 1,2, almost horizontal for 3).

\begin{table}
\caption{Stress diffusion estimates using the JS model. $\cal D\/$
is obtained from the values of $\tau_3$ (table \ref{t.1}) and
Eq.~\ref{eq:time3} with $\epsilon,K$ estimated using
$\dot\gamma_I$ and $\delta \dot\gamma$.  } \label{t.2}
\begin{center}
\begin{largetabular}{ccccccccc}
Sample& $L$ [mm] & $(\dot\gamma-\dot\gamma_I)/\dot\gamma_I$ &
$\epsilon$ & $K$ & {\cal D\/} [$m^2s^{-1}$] & $\zeta$ [nm] & $\xi$
[nm] \\\hline
1 & 0.3  & 4& $0.023$ & $0.10$ & $11.\, 10^{-15}$ & 44 & 26  \\
2 & 0.3  & 2.33 & $0.028$ & $0.12$ & $11.\, 10^{-15}$ & 43  & 26  \\
3 & 0.3  & 1.85 & $0.007$ & $0.14$ & $3.\, 10^{-15}$ & 22 & 26
 \end{largetabular}
\end{center}
\end{table}
\acknowledgments We thank J.-F.Tassin for his kind assistance and for
allowing our access to the rheometer facilities of Universit\'e du
Maine. O.R. acknowledges funding from EU COST Action P1 on Soft
Condensed Matter and from EPSRC (GR/L70455) at the beginning of this
work.

\end{document}